\documentclass[reprint,superscriptaddress,preprintnumbers,
 amsmath,amssymb,
 aps,]{revtex4-1}
\usepackage{multirow}
\usepackage{float}
\usepackage{latexsym}
\usepackage{graphicx}
\usepackage{amssymb}
\usepackage{amsmath}
\usepackage{amsfonts}
\usepackage[hidelinks]{hyperref}
\usepackage{color}
\usepackage{cool}
\usepackage{subfig}
\usepackage{dcolumn}

\begin{document}
\preprint{SLAC-PUB-17304}

\title{Interacting Dark Energy: Possible Explanation for 21-cm Absorption at Cosmic Dawn}

\author{Andr\'e A. Costa}\email{alencar@if.usp.br}
\affiliation{Instituto de F\'isica, Universidade de S\~ao Paulo, C.P. 66318, 05315-970, S\~ao Paulo, SP, Brazil}
\author{Ricardo C. G. Landim}\email{rlandim@slac.stanford.edu}\affiliation{Instituto de F\'isica, Universidade de S\~ao Paulo, C.P. 66318, 05315-970, S\~ao Paulo, SP, Brazil}\affiliation{SLAC National Accelerator Laboratory, 2575 Sand Hill Rd., Menlo Park, CA 94025 USA}
\author{Bin Wang}\email{wangb@yzu.edu.cn}\affiliation{Center for Gravitation and Cosmology, YangZhou University, Yangzhou 225009, China}
\author{Elcio Abdalla}\email{eabdalla@usp.br}
\affiliation{Instituto de F\'isica, Universidade de S\~ao Paulo, C.P. 66318, 05315-970, S\~ao Paulo, SP, Brazil}


\date{\today}

%\abstract{
\begin{abstract}
A recent observation points to an excess in the expected 21-cm brightness temperature from cosmic dawn. In this paper, we present an alternative explanation of this phenomenon, an interaction in the dark sector. Interacting dark energy models have been extensively studied recently and there is a whole variety of such in the literature. Here we particularize to a specific model in order to make explicit the effect of an interaction.
\end{abstract}

%\begin{keyword}
%Dark energy \sep Dark matter \sep 21-cm line
%\end{keyword}
%\keywords{dark energy theory, cosmological parameters from LSS}
%\pacs{98.80.Es, 98.80.Jk, 95.30.Sf}
%\arxivnumber{}
\maketitle
%\flushbottom

\section{Introduction}
Recently the EDGES collaboration observed
an absorption profile centered at 78 MHz in the sky averaged
radio spectrum \cite{Bowman:2018yin}, whose source is around redshift $z = 17$. As the first stars were formed, they emitted both Lyman-$\alpha$ photons and x-rays. This radiation penetrated the primordial hydrogen gas and changed the excitation state of the 21-cm hyperfine transition, such that photons from the cosmic microwave background (CMB) were absorbed \cite{Pritchard:2011xb}. The predicted signal at frequencies lower than $200 $ MHz  is compatible with the observed one \cite{Bowman:2018yin}. However, the observation indicates a signal with amplitude $0.5 $ K, which is more than a factor of two larger than the largest predictions \cite{Cohen:2016jbh}.

This discrepancy could be explained either increasing the temperature of the cosmic radiation or cooling the gas at the epoch of interest. In order to accomplish the latter case, it has been used a model of interaction between baryons and dark matter \cite{Barkana:2018lgd}. In this model, the interaction between baryons and dark matter could cool the hydrogen gas, giving rise to an absorption signal with the expected amplitude. On the other hand, in Ref. \cite{Munoz:2018pzp} the authors showed that the entirety of the dark matter cannot be mini-charged, but only a small amount of it could cool the baryons in the early Universe.

Although this could be a possible explanation, we would like to call attention to a phenomenon that could equally solve the problem. Along with cooling the gas or increasing the radiation temperature, another possibility is to introduce a deviation from the standard picture in the matter evolution. This can be achieved through well-known interacting dark energy models \cite{Wetterich:1994bg,Amendola:1999er,Valiviita:2008iv, He:2008si,Gavela:2009cy,Valiviita:2009nu,Martinelli:2010rt,Honorez:2010rr,Salvatelli:2013wra,Salvatelli:2014zta,DAmico:2016jbm,DiValentino:2017iww,Murgia:2016ccp,WangAbdallaetal:2016,Costa:2013sva,Costa:2014pba,Abdalla:2014cla,Costa:2016tpb} (and references therein). In those models, dark matter interacts with dark energy, implying in a different evolution as compared to the standard $\Lambda$CDM model. The interaction changes the matter evolution in the usual brightness temperature equation, which can increase the amplitude of the signal without any modification in the gas or radiation temperature.

We introduce the brightness temperature and how an interacting dark energy model can change it in section \ref{sec:temperature}. We consider a specific interacting scenario in section \ref{sec:model} and present the results in section \ref{sec:results}. Finally, section \ref{sec:conclusions} shows our conclusions.
%%%%%%%%%%%%%%%%%%%%%%%%%%%%%%%%%%%%%%%%%%%%%%%%%%%%%%%%%%%%%%%%%
\section{Brightness Temperature}
\label{sec:temperature}
Let us first consider the influence of an interacting dark energy model in the brightness temperature. We begin our discussion with the optical depth of a patch in the inter-galactic medium (IGM) in the hyperfine transition \cite{Zaldarriaga:2003du}
\begin{align}\label{tau}
\tau & = \frac{3c^3\hbar A_{10}n_{HI}}{16k_B\nu_0^2T_SH(z)} \nonumber \\ 
& \approx 8.6\times 10^{-3}x_{HI}\left[\frac{T_{CMB}(z)}{T_S}\right]\left(\frac{\Omega_bh^2}{0.02}\right) \nonumber \\
& \quad \times\left[\left(\frac{0.15}{\Omega_mh^2}\right)\left(\frac{1+z}{10}\right)\right]^{1/2} \,.
\end{align}
In this expression $c$, $\hbar$ and $k_B$ are the speed of light, the reduced Planck constant and the Boltzmann constant, respectively. $\nu_0 = 1420.4 $ MHz is the rest-frame hyperfine transition frequency and $A_{10} = 2.85 \times 10^{-15}$ s$^{-1}$ is the spontaneous emission coefficient. The number density of neutral hydrogen is given by $n_{HI}$, $T_S$ is the spin temperature of the IGM and $H(z)$ is the Hubble parameter with  $H_0 \equiv H(z=0)= 100h$  km s$^{-1}$Mpc$^{-1}$. The density parameters of matter and baryons  are represented by $\Omega_m$ and $\Omega_b$, respectively, and $x_{HI}$ is the neutral hydrogen fraction.

The brightness temperature can be calculated using the radiative transfer equation in the Rayleigh-Jeans limit, $T_b = T_{CMB}e^{-\tau} + T_S(1-e^{-\tau})$. Therefore, the intensity of the 21-cm signal relative to the CMB temperature is
\begin{align}\label{T_21}
T_{21}(z) & \approx \frac{T_S - T_{CMB}}{1+z}\tau \nonumber \\
& \approx 0.023  \times x_{HI} \left(\frac{T_S - T_{CMB}}{T_S}\right)\left(\frac{\Omega_bh^2}{0.02}\right) \nonumber \\
& \quad \times\left[\left(\frac{0.15}{\Omega_mh^2}\right)\left(\frac{1+z}{10}\right)\right]^{1/2} \text{K}\,.
\end{align}
The Hubble parameter was approximated as $H(z) \approx H_0\sqrt{\Omega_m(1+z)^3}$ in the  equations above. This is certainly true in standard cosmological models, however, in an interacting dark energy model, for instance, the dark matter is not evolving as $(1+z)^3$ anymore. We should take into account the optical depth, Eq. (\ref{tau}), with the proper Hubble parameter.
%%%%%%%%%%%%%%%%%%%%%%%%%%%%%%%%%%%%%%%%%%%%%%%%%%%%%%%%%%%%%%%%%
\section{Interacting Model}
\label{sec:model}
In interacting dark energy models the energy-momentum tensor of each component is not independently conserved. The conservation equation for dark matter (DM) and dark energy (DE) are rewritten as
\begin{equation}\label{DT_munu}
\nabla _{\mu} T^{\mu\nu}_{(i)} = Q^{\nu}_{(i)}\,,
\end{equation}
where $(i)$ represents either dark matter, $(c)$, or dark energy, $(d)$, respectively. The presence of the term $Q^{\nu}_{(i)}$ implies that there is an energy-momentum transfer between them. For instance, one can construct a phenomenological model given by the continuity equations
\begin{alignat}{2}\label{continuity}
\dot{\rho}_{c}+3\mathcal{H}\rho_{c}= & a^2Q^{0}_{c}= & +aQ\,,\nonumber \\
\dot{\rho}_{d}+3\mathcal{H}\left(1+\omega\right)\rho_{d}= & a^2Q^{0}_{d}= & -aQ \,,
\end{alignat}
where $\mathcal{H}$ is the Hubble parameter expressed in conformal time, $\mathcal{H} \equiv \dot{a}/a = a H$, and the dot represents the derivative with respect to the conformal time. The dark energy equation of state is $\omega = P_d/\rho_d$ and $Q=3H(\xi_1\rho_c + \xi_2\rho_d)$ is the phenomenological model we considered here for the energy transfer in cosmic time coordinates.

Let us assume a simpler case where the interaction is proportional to the energy density of either dark energy or dark matter only. In those cases we can solve the system of equations (\ref{continuity}) analytically. For an interaction proportional to the energy density of dark energy, $\xi_1 = 0$, we obtain \cite{He:2008tn}
\begin{align}
\label{int_DE}
\rho_{c}(z)=&~(1+z)^3 \times\left\{\frac{\xi_{2}\left[1-(1+z)^{3(\xi_{2} +\omega)}\right]\rho_{d}^0}{\xi_{2}+\omega}+\rho_{c}^0\right\} \,, \nonumber \\
\rho_{d}(z)=&~(1+z)^{3\left(1+\omega+\xi_{2}\right)}\rho_{d}^0 \,.
\end{align}
On the other hand, an interaction proportional to the energy density of dark matter, $\xi_2 = 0$, results in \cite{He:2008tn}
\begin{align}
\label{int_DM}
\rho_{c}(z)=&~(1+z)^{3-3 \xi_{1}}\rho_{c}^0  \,, \nonumber \\
\rho_{d}(z)=&~(1+z)^{3(1+\omega)} \left( \rho_{d}^{0} + \frac{\xi_{1} \rho_{c}^{0}}{\xi_{1}+\omega} \right) \nonumber \\
& ~-\frac{\xi_{1}}{\xi_{1}+\omega}(1+z)^{3(1-\xi_{1})} \rho_{c}^{0} \,.
\end{align}

\begin{figure*}[h!tb]
\subfloat[]{ \includegraphics[width=0.5\textwidth]{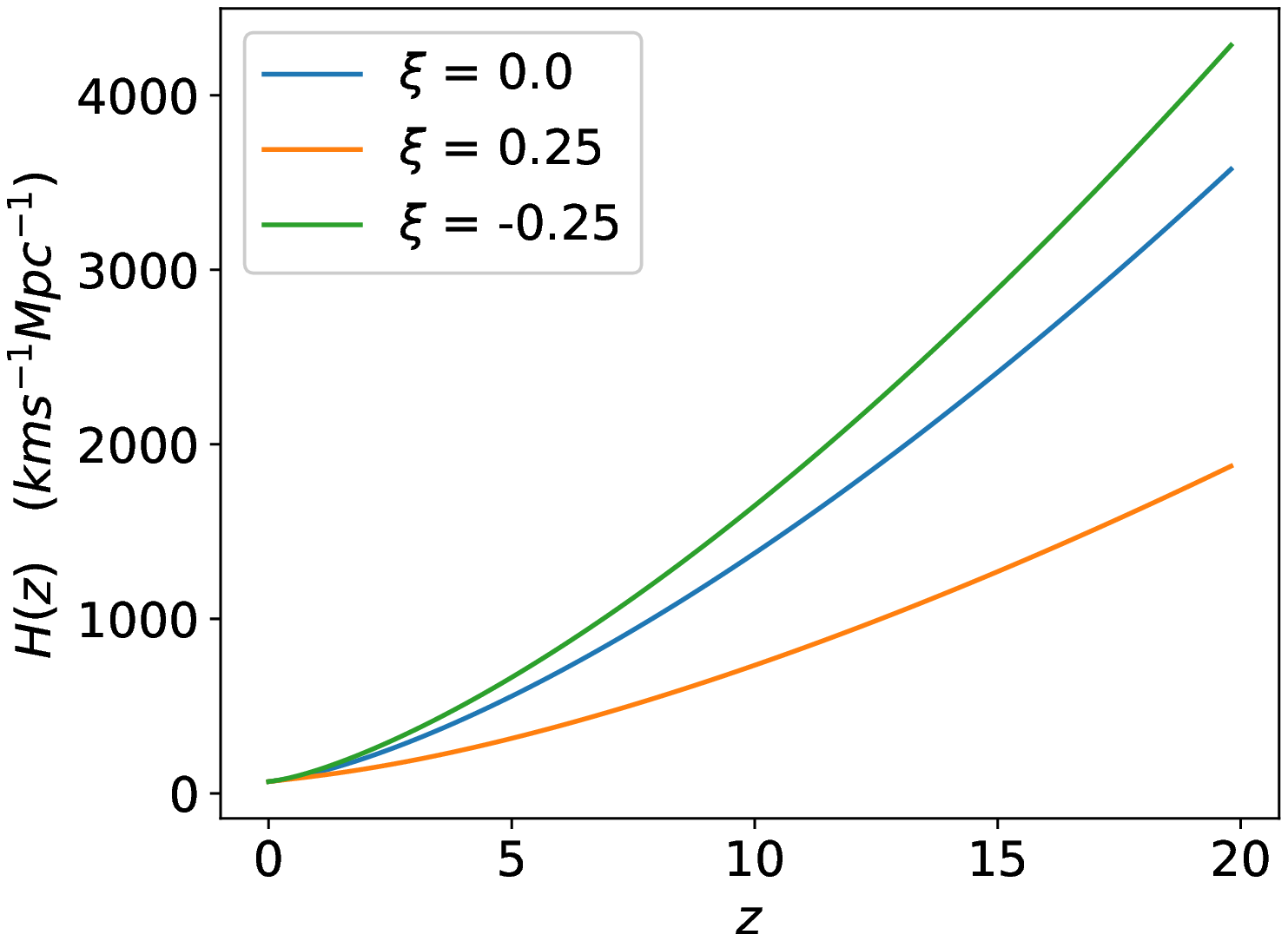}}
\subfloat[]{ \includegraphics[width=0.5\textwidth]{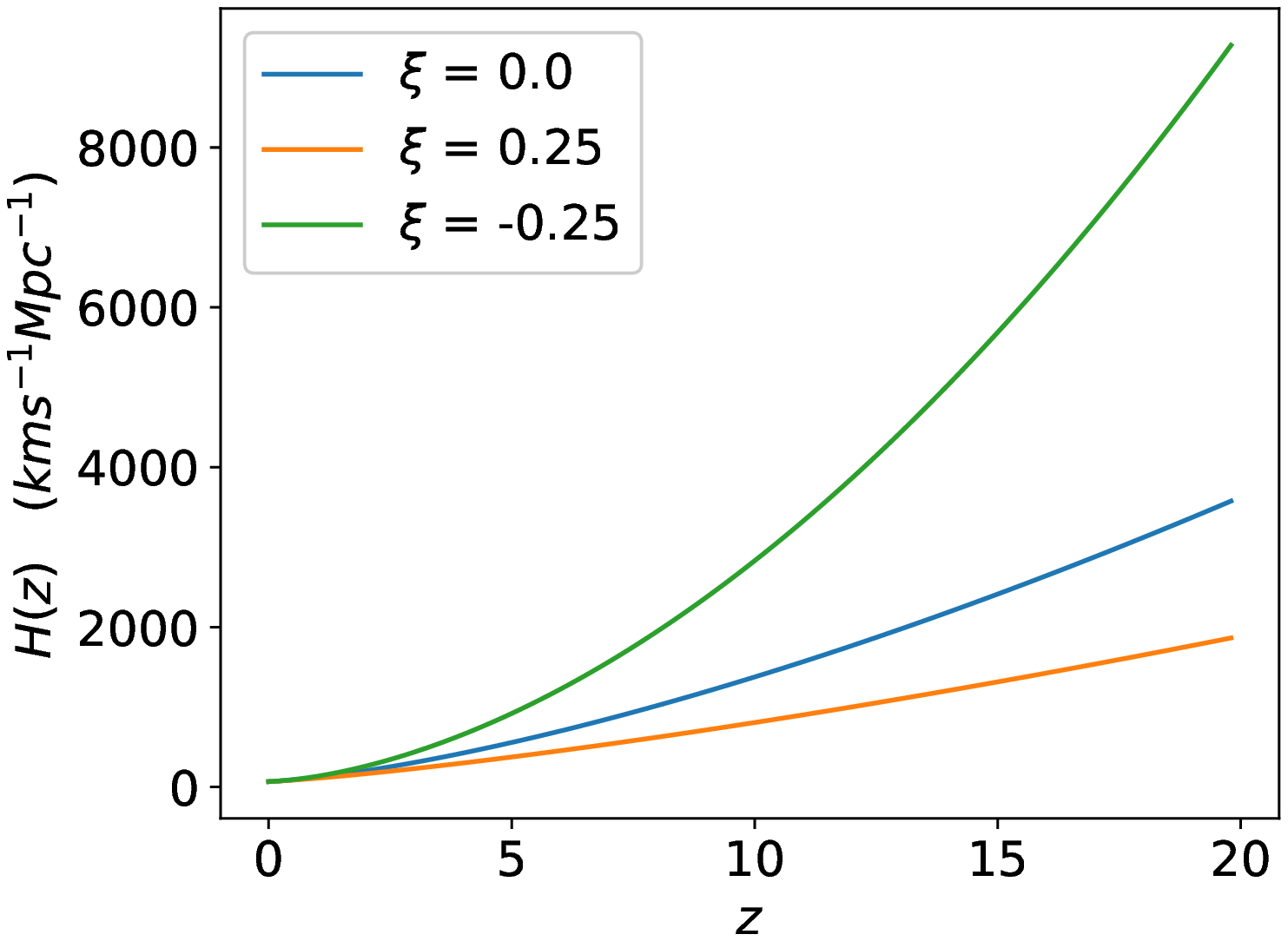}}
\caption{Hubble parameter as a function of redshift for different values of the interaction. (a) Model proportional to the energy density of DE, $Q = 3H\xi_2\rho_d$. (b) Model proportional to the energy density of DM, $Q = 3H\xi_1\rho_c$.}
\label{Hz}
\end{figure*}

Therefore, as we said before, the dark matter in an interacting model does not obey the standard $(1 + z)^3$ evolution. Taking equations (\ref{int_DM}), for instance, we see that a positive interaction will make the dark matter fall slower than the standard case, while a negative one will make it fall faster. Thus, considering the same amount of dark matter today, we would have less or more dark matter in the past, respectively. Figure \ref{Hz} shows the evolution for the Hubble parameter as a function of redshift for different values of the interaction.

These phenomenological models have instabilities with respect to curvature and dark energy perturbations \cite{Valiviita:2008iv, He:2008si,Gavela:2009cy}, depending on the values of the coupling and the dark energy equation of state. Therefore, the following  values in the parameter space should be avoided: $ i)$ for $Q \propto \rho_{d}$  ($\xi_1 = 0$),  $w < -1$ and   $\xi_2 < 0$ or  $-1 < w < 0$ and $\xi_2 > 0$; $ii)$ for $Q \propto \rho_{c}$ ($\xi_2 = 0$), $w > -1, $ for all $\xi_1 $. We note, however, that for a time-dependent equation of state this problem is less severe \cite{Xu:2011tsa}. Besides, some values of the couplings ($\xi_1$ or $\xi_2$) lead to negative energy densities for high redshifts. From Eq. (\ref{int_DE}), we see that the energy density of dark matter becomes negative in the past if $\xi_2$ is positive and $i)$ $\xi_2 + w>0$: in this case, there will always be a redshift $z$ in the past where $\rho_c < 0$ no matter the values of the other parameters; or $ii) $ $\xi_2 + w < 0 $: in this case, there will be a reshift $z$ in the past where $\rho_c < 0$ if $\Omega_{c}^{0}/\Omega_{d}^{0}<\xi_2/|\xi_2 +w|$. On the other hand, Eq. (\ref{int_DM}) shows that the energy density of dark energy reaches negative values in the past if $\xi_1 < 0$ and $i)$  $\xi_1 + w > 0$ and $ \Omega_d^{0}/ \Omega_c^{0} < | \xi_1 | / (\xi_1 + w)$ or $ii)$ $ \xi_1 + w < 0$. 
%%%%%%%%%%%%%%%%%%%%%%%%%%%%%%%%%%%%%%%%%%%%%%%%%%%%%%%%%%%%%%%%%
\section{Results}
\label{sec:results}
The conclusion from the previous section is that an interaction in the dark sector will alter the evolution of dark matter and consequently affects the Hubble parameter. This, in turn, will affect the optical depth. If the interaction makes the Hubble parameter smaller in the past, this will imply in a larger optical depth which will increase the amplitude of the 21-cm signal. Thus, even though baryons follow the standard picture, we can increase the absorption profile according to recent observations \cite{Bowman:2018yin}.

We calculated the brightness temperature for the interacting models above, Eqs. (\ref{int_DE}) and (\ref{int_DM}), according to equation (\ref{T_21}), but with a proper $H(z)$ in the optical depth equation (\ref{tau}). We calculated the Hubble parameter numerically taking into account contributions from baryons, dark matter and dark energy. Although, dark energy is sub-dominant in the redshifts of interest, we included them in our calculation for completeness, since the interaction can increase the amount of dark energy in the past. We fixed our cosmological parameters to the Planck best fit values \cite{Ade:2015xua} and considered the spin temperature to be $9.3 $ K at $z = 20$ and $5.4 $ K at $z = 15$ according to \cite{Bowman:2018yin}.

Fixing all parameters and only allowing the interaction parameter to vary, we observed that as we increase the interaction the amplitude of the 21-cm signal increases. Considering the model with interaction proportional to the energy density of dark energy, we obtained a brightness temperature $T_{21}(z) = -0.47 $ K at $z = 20$ and $T_{21}(z) = -0.56 $ K at $z = 15$ for an interaction $\xi_2 = 0.275$. Larger values for the interaction lead to negative densities and the Hubble parameter is not well behaved. Decreasing the interaction to negative values, we achieve a plateau with $T_{21}(z) = -0.11$ K at $z = 20$ and  $T_{21}(z) = -0.13$ K at $z = 15$. On the other hand, the model proportional to the energy density of dark matter has a 21-cm signal which tends to $T_{21}(z) \rightarrow -0.5 $ K at $z = 20$ and $T_{21}(z) \rightarrow -0.6$ K at $z = 15$ with an interaction $\xi_1 \rightarrow 1$ and decreasing the interaction to negative values the signal tend to zero. However, we observe that negative values for the interaction leads to a negative energy density for dark energy in the past for this model.

\begin{figure*}[h!tb]
\subfloat[]{ \includegraphics[width=0.5\textwidth]{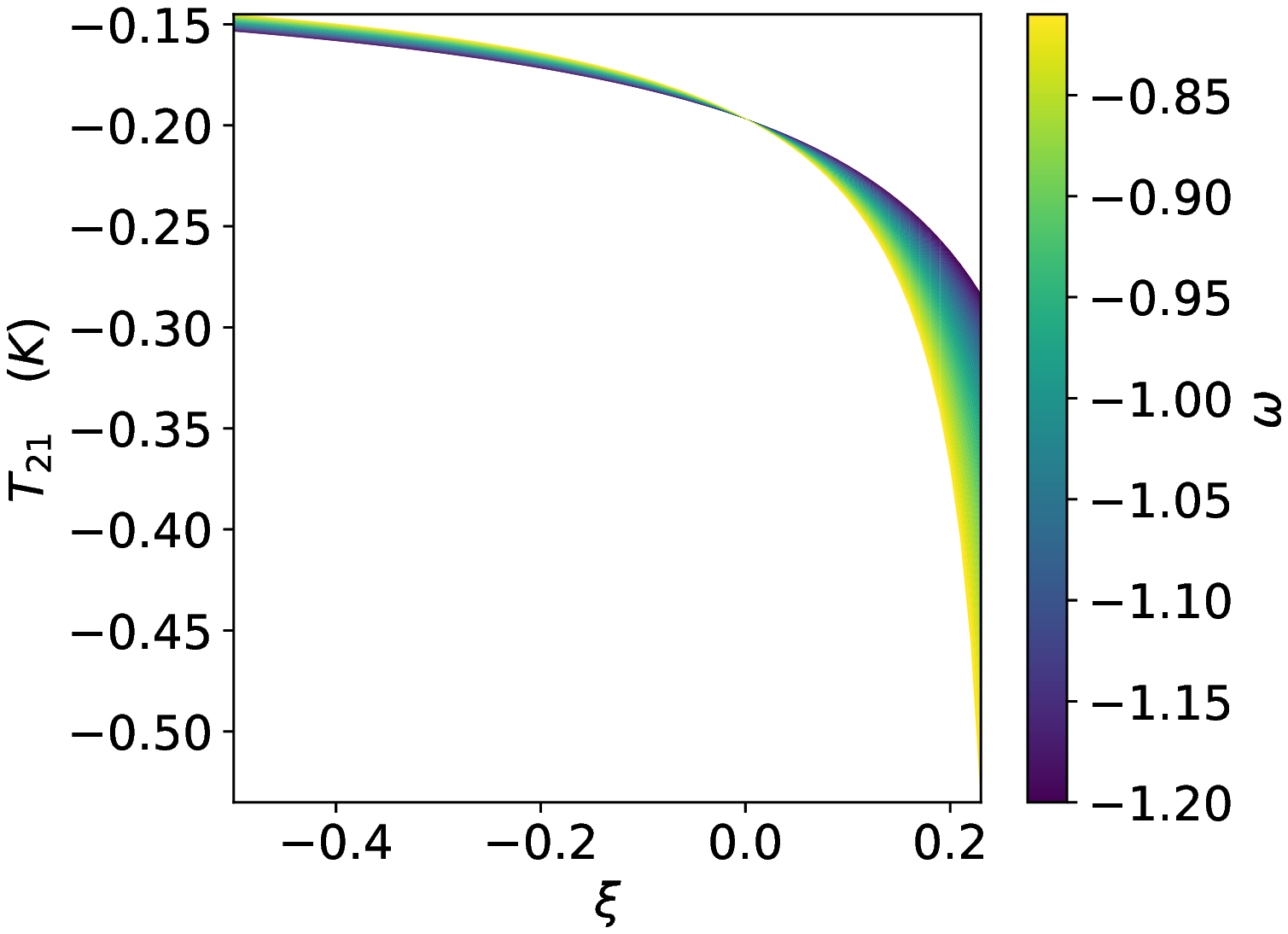}}
\subfloat[]{ \includegraphics[width=0.5\textwidth]{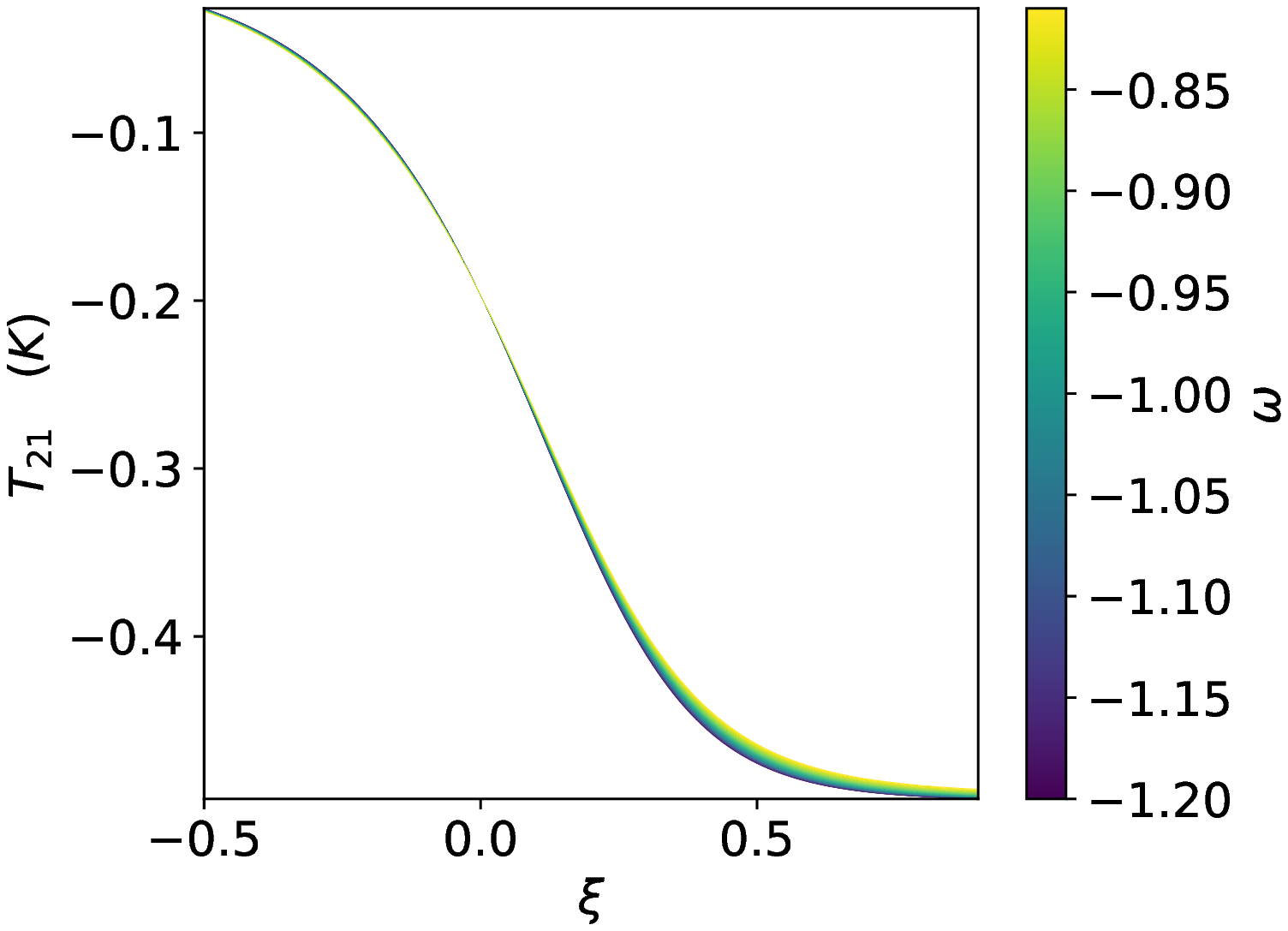}}
\caption{21-cm brightness temperature as a function of the interaction parameter and the equation of state at $z = 20$. (a) Model proportional to the energy density of DE, $Q = 3H\xi_2\rho_d$. (b) Model proportional to the energy density of DM, $Q = 3H\xi_1\rho_c$.}
\label{T21}
\end{figure*}

The above results were obtained fixing the dark energy equation of state as $\omega = -1$. Without interaction, the equation of state of dark energy has almost no effect on the brightness temperature, since the dark energy is sub-dominant. However, in the interacting case it can be important. We can see in equation (\ref{int_DE}) that the energy density of dark matter is influenced by the equation of state of dark energy. In fact, in that model, an equation of state $\omega = -1.5$ can decrease the brightness temperature from $T_{21}(z) = -0.47$ K at $z = 20$ to $T_{21}(z) = -0.28$ K in the same redshift with $\xi_2 = 0.275$. On the other direction, increasing the equation of state can produce a larger 21-cm signal, but we do not have much freedom if we want to avoid negative densities. Figure \ref{T21} shows the variation in the brightness temperature with respect to the interaction and the equation of state.

In \cite{Bowman:2018yin} the 21-cm signal, in the form of the absorption line against the CMB blackbody spectrum, was obtained where the spin temperature was lower than the CMB temperature. Our result shows that a larger positive coupling between dark sectors can present a clearer difference in the intensity of the 21-cm signal relative to the CMB temperature. This result is consistent with previous study on the reionization era, where the spin temperature was much larger than CMB temperature \cite{Xu:2017tty}. It was found, in that paper, that a larger positive interaction contributes to a stronger 21-cm power spectrum before the end of reionization \cite{Xu:2017tty}. When the interaction is proportional to the energy density of dark energy, the effect shown in the evolution of ionized fraction and the 21-cm spectrum is small and hard to be distinguished from the $\Lambda$CDM model. However, when the interaction between dark sectors is proportional to the energy density of dark matter, the reionization will be finished earlier if we fix the optical depth and we can see the difference from $\Lambda$CDM in 21-cm emission spectrum \cite{Xu:2017tty}. In the absorption spectrum profile at $78$ Mhz, we have disclosed that the interaction can show up no matter if it is proportional to the energy density of dark energy or dark matter. This strengthens our hopes to better understand the interactions between dark sectors with future 21-cm experiments.
%%%%%%%%%%%%%%%%%%%%%%%%%%%%%%%%%%%%%%%%%%%%%%%%%%%%%%%%%%%%%%%%%
\section{Conclusions}
\label{sec:conclusions}
In this letter we proposed an explanation to the recent excess in the absorption profile of the 21-cm brightness temperature. Keeping the standard behavior of baryons unaltered, we have been able to obtain an amplitude for the signal consistent with the best fit observation of $0.5$ K. This is possible because the interaction alters the Hubble expansion, especially the evolution of matter, which is dominant in those epochs.

If the Universe is described by the $\Lambda$CDM model,  EDGES result cannot be explained, although $\Lambda$CDM is quite good by comparing with other observations. This tension arises in our interacting model as well. Our interacting model is allowed by all previous observations (see \cite{Abdalla:2014cla,Costa:2016tpb}, for instance), however, using the constraints from CMB and others, we cannot find $T_{21}$ as small as the one found in EDGES. If one wants to meet the EDGES's data requirement, the interaction should be unreasonably big, which has a tension with previous constrains. On the other hand, the interaction has the effect to reduce the 21-cm value, which in turn is better than the $\Lambda$CDM model.

This tension can be used to argue that EDGES may have underestimated the brightness temperature. If it is not such small, our model can be a possible candidate. Actually, from the Wouthuysen-Field effect (see Fig.1 in \cite{Clark:2018ghm} and their Eq. (1)) the $T_{21}$ should be reduced. Whether it should be as small as measured by EDGES is a question. Finally, if one takes into account the concerns raised in \cite{Hills:2018vyr}, EDGES result does need further examination.

This measurement, if confirmed by future experiments, can also put constraints in interacting dark energy models. In particular, as the brightness temperature is not dependent on the equation of state of dark energy in the standard scenario, this kind of measurement can break the degeneracy between the interaction parameter and the equation of state.

%\acknowledgments
%\begin{acknowledgments}
A. C. acknowledges FAPESP and CAPES for the financial support under grant number 2013/26496-2, S\~ao Paulo Research Foundation (FAPESP). R. L. is supported by CNPq  under the grants 150254/2017-2 and 208206/2017-5. B. W acknowledges NNSFC. E. A. thanks FAPESP and CNPq for the financial support.
%\end{acknowledgments}

%\bibliographystyle{elsarticle-num.bst}
%\bibliographystyle{JHEP}
\bibliography{references}

\end{document}